\documentclass[
 aps,%
 11pt,%
 final,%
 notitlepage,%
 oneside,%
 twocolumn,%
 nobibnotes,%
 nofootinbib,%
 superscriptaddress,%
 noshowpacs,%
 centertags]%
 {revtex4}

\newcommand{\gdNobj}{126} 
\newcommand{\gdN}{57} 
\newcommand{\gdRh}{30}  
\newcommand{\gdVelDisp}{11}  
\newcommand{\gdMass}{$9.6\times10^{9}$}  
\newcommand{\gdLum}{$3.5\times10^8$}  
\newcommand{\gdML}{45}  

\begin{document}

 \selectlanguage{english}





\title{A List of Groups of Dwarf Galaxies in the Local Supercluster
}

\author{D.~I.~Makarov}
\email{dim@sao.ru}
\affiliation{Special Astrophysical Observatory, Russian Academy of Sciences, Nizhnii Arkhyz, 369167 Russia }

\author{R.~I.~Uklein}
\email{uklein@sao.ru}
 \affiliation{Special Astrophysical Observatory, Russian Academy of Sciences, Nizhnii Arkhyz, 369167 Russia }

\begin{abstract}
We report a list of groups consisting of dwarf galaxies only. The
sample contains  \gdNobj{} objects, mainly combined in pairs. The
most populated group contains six dwarf galaxies. The majority of
systems considered reside in the low-density regions and evolve
unaffected by massive galaxies. The characteristic sizes and
velocity dispersions of groups are \gdRh~kpc and \gdVelDisp~km/s,
respectively. They resemble the associations of dwarf galaxies,
but are more compact. On the whole, groups and associations form a
continuous sequence. Alike the associations, our groups possess
high mass-to-luminosity ratios, what is indicative of a large
amount of dark matter present in these systems.
\end{abstract}

\maketitle

\section{INTRODUCTION}

Contemporary mass surveys, such as 2dF~\cite{colless01:Makarov},
HIPASS~\cite{zwaan03:Makarov}, 6dF~\cite{jones04:Makarov},
ALFALFA~\cite{alfalfa:Makarov} and SDSS~\cite{sdss7:Makarov} have
substantially enriched  our knowledge of redshifts in the near
Universe. The number of measured velocities within the Local
Supercluster has increased by a factor of three to four over the
past decade. Owing to all-sky surveys, redshifts have become known
not only for the giant, but also for a large number of dwarf
galaxies.

In their recent series of papers, Makarov and
Karachentsev~\cite{karachentsev08_pairs:Makarov,makarov09_triplets:Makarov,makarov11_groups:Makarov}
have studied the distribution and properties of multiple systems
of galaxies on the scale of the Local Supercluster.
In~\cite{karachentsev08_pairs:Makarov} a large number of pairs
consisting exclusively of dwarf galaxies  was  brought to notice.
Extremely metal-poor galaxies can be found among such systems,
like, e.g., the well-known system I\,Zw\,18 and the pair of
galaxies HS\,0822+3542a and
SAO\,0822+3545~\cite{chengalur06:Makarov}. Most of the multiple
dwarf galaxies contain young stellar population, which is
reflected in the color and morphology of these systems. Radio
observations of such objects have shown that they contain large
volumes of neutral hydrogen~\cite{ekta06:Makarov}.

Tully et al.~\cite{tully06:Makarov} used high-precision
photometric distances to study the three-dimensional distribution
of nearby galaxies on the 3~Mpc scale and identified systems of
dwarf galaxies in the neighborhood of the Local Group. Such
structures, which were called the associations of dwarf galaxies,
have the mass-to-luminosity ratios in the
 \mbox{100--1000 $M_{\sun} /L_{\sun}$} interval and contain large amounts of
dark matter. The discovery of pairs of dwarf galaxies on the scale
of the Local Supercluster is indicative of a wide occurrence of
such systems in the Universe.

\section{GROUPS IN THE LOCAL SUPERCLUSTER}

This paper continues a series of papers by Makarov and
Karachentsev aimed at the study of multiple systems on the scale
of
40~Mpc~\cite{karachentsev08_pairs:Makarov,makarov09_triplets:Makarov,makarov11_groups:Makarov}.
These papers provide a detailed description of the technique of
identification of groups and analysis of the sample obtained. Here
we only briefly describe the necessary details of catalog
creation.  Radial velocities, magnitudes, morphological types, and
other galaxy parameters were adopted from the HyperLEDA and NED
databases. Both databases contain a large amount of false data,
which appear due to the automatic procedures of data reduction.
The most common types of pollution are: confusion of the
coordinates and velocities of galaxies closely located in the sky;
objects with false line-of-sight velocities obtained from the mass
2dF-type sky surveys; apparent magnitudes from the Sloan Digital
Sky Survey (SDSS) survey refer to individual regions of extended
galaxies. We corrected these and some other errors as far as
possible. Visual control of galaxy parameters was a very important
and most time-consuming stage when working with the catalogs of
groups~\cite{karachentsev08_pairs:Makarov,makarov09_triplets:Makarov,makarov11_groups:Makarov}.
As a result, we obtained a sample of 10\,914 galaxies with
line-of-sight velocities in the reference frame of the Local Group
\mbox{$V_{\rm LG} < 3500$~km/s} located at the galactic latitudes
of \mbox{$|b|
> 15\degr$}. A sample of such a depth comprises the entire Local
Supercluster with all its neighborhoods.

The clustering algorithm for multiple \linebreak
systems~\mbox{\cite{karachentsev08_pairs:Makarov,makarov09_triplets:Makarov,makarov11_groups:Makarov}}
is based on the natural requirement that the total energy of the
physical pair should be negative. At the first stage we computed
the boundedness  criteria for all galaxy pairs in the sample: the
total energy of the system should satisfy the inequality
\mbox{$\Delta V^2 \Delta R < 2 G M$} and the galaxies should
reside inside the zero-velocity sphere \mbox{$\pi H_0^2 \Delta R^3
< 8 G M$}, where $\Delta V$ and $\Delta R$ are the velocity
difference and the difference of projected distances in a pair of
galaxies, and $M$ is their total mass. We combined the pairs
selected by these criteria into groups and repeated the process
while there was at least one pair meeting the above criteria. The
algorithm uses only the information on the coordinates, redshifts,
and magnitudes of objects. We determine the distances to galaxies
using the Hubble law with \mbox{$H_0 = 73$~km/s/Mpc}. The masses
were estimated by the near-infrared integrated $Ks$-band
magnitudes of galaxies assuming that all galaxies have equal
mass-to-luminosity ratios.

We adopt most of the photometric data from the 2MASS all-sky
survey~\cite{jarret00:Makarov,jarret03:Makarov}. In the absence of
$K$-band estimates we converted the optical  ($B$, $V$, $R$, $I$)
and near-infrared ($J$, $H$) magnitudes into the $K$-band
magnitudes, as described in the series of
papers~\mbox{\cite{karachentsev08_pairs:Makarov,makarov09_triplets:Makarov,makarov11_groups:Makarov}}.

As a result of clustering,  5926 galaxies have been combined into
1082 systems consisting of two and more members. Makarov and
Karachentsev~\cite{makarov11_groups:Makarov} showed that the
median velocity dispersions and harmonic radii of groups with
$n\ge4$ members in the Local Supercluster are equal to
\mbox{$\sigma_V=74$~km/s} and $204$~kpc, respectively; the mean
crossing time is about 2.2~Gyr; the typical mass of a group is
\mbox{$M_p = 2.3\times10^{12}~M_{\sun}$} and the $K$-band
mass-to-luminosity ratio is \mbox{$M/L = 22~M_{\sun}/L_{\sun}$}.

\section{GROUPS OF DWARF GALAXIES}

Karachentsev and Makarov~\cite{karachentsev08_pairs:Makarov}
pointed at the existence of an unexpectedly large number of pairs
consisting of dwarf galaxies. Although such systems have been
already known for a while, including the most famous of them---the
pair of galaxies I\,Zw\,18\,A and
I\,Zw\,18\,C~\cite{izw18a89:Makarov}, the groups of dwarf galaxies
have until recently remained outside the focus of attention.

\begin{figure}[t]
\setcaptionmargin{0mm}
\includegraphics[scale=0.65]{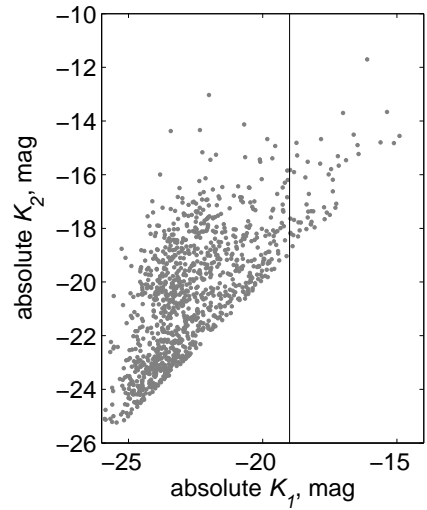}
\caption{Dependence of the absolute magnitude  of the second
brightest member $K_2$  on the absolute magnitude  of the
brightest member $K_1$. The groups of dwarf galaxies are located
to the right from the solid line \mbox{($M_K>-19$ mag)}. This plot
corresponds to the right-hand panel of Fig.~7
from~\cite{karachentsev08_pairs:Makarov}. } \label{f:k1k2:Makarov}
\end{figure}

Figure~\ref{f:k1k2:Makarov} demonstrates the distribution of
luminosities of the brightest and second brightest group members
for the groups from~\cite{karachentsev08_pairs:Makarov}. The
systems that have caught our attention are located to the right
from the vertical line. We compiled a list of groups of dwarf
galaxies based on the catalogs of groups in the Local
Supercluster~\cite{karachentsev08_pairs:Makarov,makarov09_triplets:Makarov,makarov11_groups:Makarov}.
The sample of galaxies from these papers was slightly modified. We
added new and updated the available line-of-sight velocities and
magnitudes of galaxies using the published data. The changes made
in the process of refinement and correction of the HyperLEDA data
were taken into account. We also updated the data on groups of
galaxies in the Local Supercluster using the original clustering
algorithm. The list of dwarf galaxies now includes systems, in
which the brightest galaxy has the $K$-band absolute magnitude
fainter than \mbox{$M_K=-19$ mag}, while these groups are not the
substructures of brighter formations. This allowed us to mainly
select dwarf irregular galaxies with a small number of late-type
galaxies (Sdm--Sm) among the brightest objects.

We cleaned the resulting sample from false systems. Such
formations may appear in the regions with high negative peculiar
velocities near the clusters of galaxies. The use of the Hubble
law for determining the distances to them may result in a
substantial underestimation of total luminosities of galaxies.

We visually searched for possible group members without known
line-of-sight velocities around the selected systems in the region
of approximately \mbox{$1\degr \times 1\degr$}. The main criterion
of visual selection was the morphological consistency between the
candidates and known members and the agreement of their redshifts.
We performed our search only within the SDSS coverage region. Our
list also incorporated multiple dwarf systems, that we identified
in the process of visual inspection of images within the
verification of galaxy data reliability in the HyperLEDA database.
This was not a systematic search and it affected only the galaxies
that have caught our attention for some reason.

Note that a visual search for multiple dwarf systems on the SDSS
images revealed more than 20 candidates, which subsequently failed
to pass the isolation criteria. These close dwarf systems proved
to be subsystems located around bright galaxies or inside the more
massive groups.

We present the list of groups of dwarf galaxies in
Table~\ref{catalog:Makarov}. The columns of the table give:
(1)~the name of the group in the list; (2)~the designation of the
component based on the number of a given galaxy in the order of
increasing right ascension; (3)~the name of the galaxy according
to common catalogs; (4)~J2000 coordinates of the galaxy;
(5)~line-of-sight velocity $V_{LG}$ with respect to the center of
the Local Group in accordance with~\cite{apex1996:Makarov};
(6)~apparent $B$-band magnitude, which was estimated from the SDSS
$g$ and $r$-band photometry (see Table~1
from~\cite{Jester2005:Makarov}) or adopted from the HyperLEDA
database~\cite{HL:Makarov}; (7)~the absolute $B$-band  magnitude;
(8)~the main disturber (MD) and index of isolation ($II$) of the
group.

\onecolumngrid
\setcaptionwidth{\linewidth}%

 \begin{longtable*}{c|c|l|cc|cr@{$\,\pm\,$}ll|c|c|l|r}
\caption{The list of dwarf galaxy groups}\label{catalog:Makarov}\\
\hline
Group && \multicolumn{1}{c|}{Galaxy name} & R.A. & Dec & & \multicolumn{3}{c|}{$V_{\rm LG}$,} & $m_B$,  & $M_B$, & \multicolumn{1}{c|}{MD} & \multicolumn{1}{c}{$II$} \\
&&&\multicolumn{2}{c|}{J2000} & & \multicolumn{3}{c|}{km/s} & mag & mag &&\\
(1) & (2)  & \multicolumn{1}{c|}{(3)} & \multicolumn{2}{c|}{(4)} & &\multicolumn{3}{c|}{(5)} & (6) & (7) & \multicolumn{1}{c|}{(8)} & \multicolumn{1}{c}{(9)}\\
\endfirsthead

\caption{(Contd.) }\\
\hline
Group && \multicolumn{1}{c|}{Galaxy name} & R.A. & Dec & & \multicolumn{3}{c|}{$V_{\rm LG}$,} & $m_B$,  & $M_B$, & \multicolumn{1}{c|}{MD} & \multicolumn{1}{c}{$II$} \\
&&&\multicolumn{2}{c|}{J2000}& & \multicolumn{3}{c|}{km/s} & mag & mag &&\\
(1) & (2)  & \multicolumn{1}{c|}{(3)} & \multicolumn{2}{c|}{(4)} & & \multicolumn{3}{c|}{(5)} & (6) & (7) & \multicolumn{1}{c|}{(8)} & \multicolumn{1}{c}{(9)}\\
\hline
\endhead

\hline
\endfoot

\hline
\endlastfoot

\hline
J0130+02   & A & UGC\,1075                    & 01\,30\,02.5&$+$02\,51\,09 && 2227 &   6 && 16.60 & $-$15.92 & NGC\,488   &  1.29  \\
           & B & LSBC\,F828-01                & 01\,30\,29.0&$+$02\,49\,55 && 2240 &  10 && 17.00 & $-$15.53 &            &        \\[2.0mm]

J0310$-$41 & A & LCRS\,B030900.1$-$415914     & 03\,10\,49.7&$-$41\,47\,57 && 1253 &  74 && 16.03 & $-$15.20 & NGC\,1399  &  1.75  \\
           & B & LCRS\,B030909.4$-$415056     & 03\,10\,59.2&$-$41\,39\,40 && 1186 &  26 &&       &          &            &        \\[2.0mm]

J0453$-$61 & A & ESO\,119-016                 & 04\,51\,29.2&$-$61\,39\,03 &&  739 &  10 && 14.89 & $-$15.24 & NGC\,1796  &  1.19  \\
           & B & SGC\,0454.2$-$6138           & 04\,54\,55.4&$-$61\,33\,53 &&  745 &   9 && 16.09 & $-$14.05 &            &        \\[2.0mm]

J0532$-$25 & A & ESO\,487-017                 & 05\,30\,29.0&$-$24\,52\,35 && 1661 &  13 && 16.24 & $-$15.76 &            &        \\
           & B & ESO\,487-020                 & 05\,32\,23.8&$-$25\,13\,55 && 1750 &  74 && 16.14 & $-$15.92 & NGC\,1964  &  5.79  \\
           & C & AM\,0530$-$245               & 05\,32\,46.5&$-$24\,55\,33 && 1684 &  74 && 16.67 & $-$15.31 &            &        \\[2.0mm]

J0700$-$04 & A & HIZSS\,003\,A                & 07\,00\,28.2&$-$04\,12\,26 &&  109 &  11 && 18.00 & $-$12.29 & LG         &  5.89  \\
           & B & HIZSS\,003\,B                & 07\,00\,24.6&$-$04\,13\,13 &&  143 &  11 &&       &          &            &        \\[2.0mm]

J0714+44   & A & UGC\,3698                    & 07\,09\,18.7&$+$44\,22\,48 &&  465 &   5 && 15.16 & $-$14.35 & M\,81      & 22.90  \\
           & B & NGC\,2337                    & 07\,10\,13.6&$+$44\,27\,26 &&  477 &   5 && 13.10 & $-$16.39 &            &        \\
           & C & UGC\,3817                    & 07\,22\,44.5&$+$45\,06\,31 &&  478 &   2 && 15.96 & $-$13.63 &            &        \\[2.0mm]

J0723+36   & A & SDSS\,J072313.46+362213.0    & 07\,23\,13.5&$+$36\,22\,13 &&  967 &   1 && 19.31 & $-$11.54 &            &        \\
           & B & SDSS\,J072301.42+362117.1    & 07\,23\,01.4&$+$36\,21\,17 &&  914 &   1 && 17.01 & $-$13.72 &            &        \\
           & C & SDSS\,J072320.56+362440.9    & 07\,23\,20.6&$+$36\,24\,41 &&  935 &   1 && 21.59 & $-$9.19  &            &        \\[2.0mm]

J0742+16   & A & UGC\,3974                    & 07\,41\,55.4&$+$16\,48\,09 &&  162 &   5 && 13.62 & $-$13.24 & M\,81      & 49.14  \\
           & B & CGCG\,087-033                & 07\,42\,32.0&$+$16\,33\,40 &&  168 &   5 && 15.43 & $-$11.49 &            &        \\[2.0mm]

J0747+51   & A & MCG\,+09-13-052              & 07\,46\,57.0&$+$51\,17\,47 &&  510 &   5 && 16.66 & $-$12.85 & NGC\,2500  & 12.11  \\
           & B & KUG\,0743+513                & 07\,47\,32.0&$+$51\,11\,29 &&  503 &   5 && 15.14 & $-$14.32 &            &        \\[2.0mm]

J0817+24   & A & LCSB\,S1123P                 & 08\,17\,15.9&$+$24\,53\,57 && 1832 &   5 && 17.11 & $-$15.17 & IC\,2267    &  8.42  \\
           & B & KUG\,0814+251                & 08\,17\,21.0&$+$24\,57\,46 && 2076 &  5 &$^{\dag}$& 17.12 & $-$15.43 &            &        \\[2.0mm]

J0821$-$00 & A & UGC\,4358                    & 08\,21\,26.0&$-$00\,25\,08 && 1606 &   6 && 15.96 & $-$15.94 & UGC\,4254  & 74.24  \\
           & B & 6dF\,J0821428$-$002601       & 08\,21\,42.8&$-$00\,26\,01 && 1612 &  74 && 16.48 & $-$15.43 &            &        \\[2.0mm]

J0825+35   & A & HS\,0822+3542a               & 08\,25\,55.5&$+$35\,32\,32 &&  698 &   3 && 16.57 & $-$13.53 & NGC\,2683  & 43.55  \\
           & B & SAO\,0822+3545               & 08\,26\,05.6&$+$35\,35\,26 &&  712 &   2 && 18.34 & $-$11.80 &            &        \\[2.0mm]

J0852+13   & A & SDSS\,J085233.76+135028.4    & 08\,52\,33.8&$+$13\,50\,28 && 1360 &   3 && 17.15 & $-$14.34 &            &        \\
           & B & SDSS\,J085240.94+135157.0    & 08\,52\,40.9&$+$13\,51\,57 && 1390 &  22 && 19.56 & $-$11.98 &            &        \\[2.0mm]

J0859+39   & A & UGC\,4704                    & 08\,58\,59.0&$+$39\,12\,40 &&  581 &   6 && 15.51 & $-$14.12 & NGC\,2683  & 18.29  \\
           & B & SDSS\,J085947.03+392302.6    & 08\,59\,46.9&$+$39\,23\,06 &&  573 &  34 && 17.23 & $-$12.37 &            &        \\[2.0mm]

J0911+42   & A & SDSS\,J091108.40+423922.1    & 09\,11\,08.4&$+$42\,39\,22 && 1498 &  13 && 16.02 & $-$15.61 & NGC\,2798  & 14.93  \\
           & B & SDSS\,J091110.62+423801.4    & 09\,11\,10.6&$+$42\,38\,01 &&      &     && 18.81 &          &            &        \\[2.0mm]

J0915+48   & A & UGC\,4868                    & 09\,14\,51.8&$+$48\,35\,37 && 2822 &   5 && 17.43 & $-$15.58 & NGC\,2856  &  4.61  \\
           & B & UGC\,4874                    & 09\,15\,16.3&$+$48\,40\,03 && 2821 &   5 && 17.60 & $-$15.41 &            &        \\
           & C & SDSS\,J091552.07+484119.5    & 09\,15\,52.1&$+$48\,41\,20 && 2809 &  17 && 18.08 & $-$14.94 &            &        \\[2.0mm]

J0934+55   & A & I\,Zw\,18\,C                 & 09\,33\,59.7&$+$55\,14\,45 &&  821 &   5 && 19.73 & $-$10.66 & NGC\,2841  &  8.34  \\
           & B & I\,Zw\,18\,A                 & 09\,34\,02.0&$+$55\,14\,28 &&  837 &   4 && 16.48 & $-$13.95 &            &        \\[2.0mm]

J0950+31   & A & UGC\,5272b                   & 09\,50\,19.5&$+$31\,27\,22 &&  479 &   5 && 17.82 & $-$11.35 & NGC\,2903  & 12.64  \\
           & B & UGC\,5272                    & 09\,50\,22.4&$+$31\,29\,16 &&  453 &   4 && 14.46 & $-$14.59 &            &        \\[2.0mm]

J0959+41   & A & KUG\,0956+420                & 09\,59\,30.0&$+$41\,46\,01 && 1704 &  36 && 16.43 & $-$15.47 & NGC\,2964  & 99.46  \\
           & B & KUG\,0956+419                & 09\,59\,45.0&$+$41\,40\,37 && 1664 &   3 &$^{\dag}$& 16.61 & $-$15.23 &            &        \\[2.0mm]

J1016+37   & A & UGC\,5540                    & 10\,16\,21.9&$+$37\,46\,47 && 1138 &   4 && 14.60 & $-$16.42 & NGC\,3245  & 66.28  \\
           & B & KUG\,1013+381                & 10\,16\,24.5&$+$37\,54\,46 && 1150 &   3 && 15.97 & $-$15.07 &            &        \\[2.0mm]

J1040$-$09 & A & 6dF\,J1039304$-$094609       & 10\,39\,30.4&$-$09\,46\,09 && 2177 &  74 && 16.68 & $-$15.85 & NGC\,3375  &  7.86  \\
           & B & 6dF\,J1040118$-$095641       & 10\,40\,11.8&$-$09\,56\,40 && 2218 &  74 && 16.48 & $-$16.08 &            &        \\[2.0mm]

J1052+00   & A & MGC\,0013223                 & 10\,52\,40.6&$-$00\,01\,17 && 1569 &  75 && 17.59 & $-$14.26 & UGC\,5922  &  8.11  \\
           & B & CGCG\,010-041                & 10\,52\,48.6&$+$00\,02\,04 && 1607 &   5 && 15.62 & $-$16.28 &            &        \\[2.0mm]

J1053+02   & A & LSBC\,L1-137A                & 10\,53\,03.1&$+$02\,29\,37 &&  860 &   5 && 17.59 & $-$12.93 & NGC\,3379  &  5.35  \\
           & B & LSBC\,L1-137                 & 10\,53\,18.6&$+$02\,37\,34 &&  851 &  10 && 15.80 & $-$14.69 &            &        \\[2.0mm]

J1101+30   & A & BTS\,028                     & 11\,01\,32.4&$+$30\,35\,16 && 1708 &  75 && 17.56 & $-$14.38 & NGC\,3430  &  6.38  \\
           & B & BTS\,029                     & 11\,01\,38.9&$+$30\,36\,29 && 1626 &   5 && 18.32 & $-$13.52 &            &        \\[2.0mm]

J1110+40   & A & KUG\,1107+403                & 11\,10\,25.2&$+$40\,03\,11 && 2943 &  30 && 16.11 & $-$16.98 & NGC\,3665  & 85.47  \\
           & B & SDSS\,J111026.28+400117.4    & 11\,10\,26.3&$+$40\,01\,17 &&      &     && 19.06 &          &            &        \\[2.0mm]

J1113+53   & A & UGC\,06251                   & 11\,13\,26.1&$+$53\,35\,42 &&  999 &   5 && 16.33 & $-$14.42 & NGC\,3992  &  2.14  \\
           & B & SDSS\,J111343.60+533848.3    & 11\,13\,43.6&$+$53\,38\,48 &&  985 &   5 && 18.04 & $-$12.68 &            &        \\[2.0mm]

J1131$-$35 & A & 6dF\,J1131390$-$352255       & 11\,31\,38.9&$-$35\,22\,56 && 2396 &  48 && 16.52 & $-$16.36 & NGC\,3742  &  3.50  \\
           & B & PGC\,649656                  & 11\,31\,39.6&$-$35\,22\,42 &&      &      && 17.35 &          &            &        \\[2.0mm]

J1134+48   & A & SDSS\,J113342.71+482004.9    & 11\,33\,42.7&$+$48\,20\,05 && 3094 &  20 && 17.69 & $-$15.53 & NGC\,3811  &  5.94  \\
           & B & SDSS\,J113403.75+482834.4    & 11\,34\,03.9&$+$48\,28\,37 && 3107 &  29 && 18.15 & $-$15.09 &            &        \\[2.0mm]

J1141+32   & A & KUG\,1138+327                & 11\,41\,07.4&$+$32\,25\,37 && 1704 &  70 && 15.86 & $-$16.08 & IC\,2957    &  5.64  \\
           & B & MRK\,0746                    & 11\,41\,29.9&$+$32\,20\,59 && 1684 &  50 && 15.68 & $-$16.24 &            &        \\
           & C & SDSS\,J114136.70+321651.5    & 11\,41\,36.7&$+$32\,16\,52 && 1737 &   2 && 17.00 & $-$14.98 &            &        \\[2.0mm]

J1146+58   & A & SBS\,1143+588                & 11\,45\,58.7&$+$58\,32\,07 && 1519 &  37 && 15.75 & $-$15.89 & NGC\,4036  &  2.00  \\
           & B & SDSS\,J114603.39+583621.8    & 11\,46\,03.4&$+$58\,36\,22 && 1518 &  42 && 17.81 & $-$13.83 &            &        \\[2.0mm]

J1150$-$00 & A & UM\,456A                     & 11\,50\,34.0&$-$00\,32\,16 && 1645 &  16 && 17.07 & $-$14.81 & NGC\,4472  & 10.39  \\
           & B & UM\,456                      & 11\,50\,36.2&$-$00\,34\,02 && 1574 &   5 && 16.43 & $-$15.35 &            &        \\[2.0mm]

J1152$-$02 & A & UM\,461                      & 11\,51\,33.4&$-$02\,22\,22 &&  866 &   9 && 14.71 & $-$15.74 & NGC\,4472  &  6.02  \\
           & B & UGC\,6850                    & 11\,52\,37.4&$-$02\,28\,10 &&  860 &   6 && 14.72 & $-$15.72 &            &        \\[2.0mm]

J1154$-$03 & A & SDSS\,J115348.29$-$031306.5  & 11\,53\,48.3&$-$03\,13\,06 && 1243 &  12 && 18.00 & $-$13.25 &           & \\
           & B & CGCG\,012-113                & 11\,54\,07.6&$-$03\,40\,56 && 1223 &  27 && 16.02 & $-$15.20 & NGC\,4030  &  2.30  \\
           & C & SDSS\,J115503.67$-$033012.4  & 11\,55\,03.7&$-$03\,30\,12 && 1216 &  69 && 18.02 & $-$13.20 &            &        \\[2mm]

J1157+02   & A & SDSS\,J115725.14+021115.9    & 11\,57\,25.1&$+$02\,11\,16 &&  839 &   5 && 17.95 & $-$12.45 & NGC\,4472  &  2.43  \\
           & B & SDSS\,J115735.27+021004.0    & 11\,57\,35.3&$+$02\,10\,04 &&  796 &  32 && 16.48 & $-$13.81 &            &        \\[2.0mm]

J1158+31   & A & KDG\,083                     & 11\,56\,14.5&$+$31\,18\,16 &&  617 &   5 && 16.65 & $-$13.06 & NGC\,4278  &  2.70  \\
           & B & KUG\,1157+315                & 12\,00\,16.2&$+$31\,13\,30 &&  593 &  28 && 15.25 & $-$14.36 &            &        \\[2.0mm]

J1216+52   & A & CGCG\,269-049                & 12\,15\,46.8&$+$52\,23\,17 &&  245 &  15 && 15.27 & $-$12.47 & NGC\,3031  &  5.51  \\
           & B & UGC\,07298                   & 12\,16\,30.1&$+$52\,13\,39 &&  254 &   5 && 16.64 & $-$11.15 &            &        \\[2.0mm]

J1221+38   & A & KUG\,1218+387                & 12\,20\,54.9&$+$38\,25\,49 &&  623 &  46 && 15.57 & $-$14.16 & NGC\,4490  &  1.95  \\
           & B & KDG\,105                     & 12\,21\,43.0&$+$37\,59\,14 &&  582 &   5 && 17.57 & $-$11.98 &            &        \\[2.0mm]

J1224+28   & A & [KK98]\,138                  & 12\,21\,58.4&$+$28\,14\,34 &&  417 &   8 && 18.88 & $-$10.00 & NGC\,4278  &  2.28  \\
           & B & [KK98]\,144                  & 12\,25\,29.1&$+$28\,28\,57 &&  453 &   2 && 18.18 & $-$10.87 &            &        \\[2.0mm]

J1225+61   & A & MCG\,+10-18-044              & 12\,24\,53.8&$+$61\,03\,49 &&  833 &   5 && 15.85 & $-$14.50 & NGC\,3992  &  9.87  \\
           & B & SBS\,1222+614                & 12\,25\,05.4&$+$61\,09\,11 &&  832 &   5 && 14.86 & $-$15.49 &            &        \\[2.0mm]

J1226$-$01 & A & UGC\,7531                    & 12\,26\,11.8&$-$01\,18\,17 && 1858 &  12 && 15.27 & $-$16.86 & NGC\,4527  &  5.13  \\
           & B & UM\,501                      & 12\,26\,22.7&$-$01\,15\,12 && 1861 &  14 && 16.41 & $-$15.72 &            &        \\[2.0mm]

J1228+22   & A & UGC\,7584                    & 12\,28\,02.8&$+$22\,35\,16 &&  545 &   4 && 16.20 & $-$13.25 & NGC\,4278  &  3.19  \\
           & B & KKH\,80                      & 12\,28\,05.0&$+$22\,17\,27 &&  543 &   5 && 17.00 & $-$12.44 &            &        \\
           & C & NGC\,4455                    & 12\,28\,44.1&$+$22\,49\,14 &&  581 &   5 && 13.05 & $-$16.55 &            &        \\[2.0mm]

J1244+62   & A & MCG\,+11-16-003              & 12\,43\,59.9&$+$62\,19\,60 && 2739 &  77 && 16.46 & $-$16.48 & NGC\,4521  &  7.41  \\
           & B & MCG\,+11-16-005              & 12\,44\,12.0&$+$62\,14\,51 && 2750 &   9 && 16.10 & $-$16.85 &            &        \\
           & C & SDSS\,J124411.92+621021.5    & 12\,44\,12.1&$+$62\,10\,19 && 2681 &  12 && 18.00 & $-$14.88 &            &        \\
           & D & SDSS\,J124418.07+621007.7    & 12\,44\,18.0&$+$62\,10\,07 &&      &     && 18.32 &          &            &        \\
           & E & SDSS\,J124423.23+620305.5    & 12\,44\,23.2&$+$62\,03\,06 && 2660 &  72 && 17.86 & $-$15.01 &            &        \\[2.0mm]

J1301$-$01 & A & CGCG\,1258.5$-$0142S         & 13\,01\,00.7&$-$01\,58\,34 && 1302 &   5 && 17.04 & $-$14.31 & NGC\,4699  &  3.11  \\
           & B & UGC\,8127                    & 13\,01\,03.7&$-$01\,57\,12 && 1297 &  33 && 15.67 & $-$15.67 &            &        \\[2.0mm]

J1303$-$17 & A & UGCA\,319                    & 13\,02\,14.4&$-$17\,14\,15 &&  548 &   8 && 15.08 & $-$14.65 & NGC\,5068  & 10.32  \\
           & B & UGCA\,320                    & 13\,03\,16.7&$-$17\,25\,23 &&  546 &   5 && 13.40 & $-$16.31 &            &        \\[2.0mm]

J1304$-$02 & A & LCRS\,B130157.2$-$024313     & 13\,04\,31.8&$-$02\,59\,17 && 1148 &  34 && 16.27 & $-$14.82 & NGC\,4699  &  2.26  \\
           & B & HIPASS\,J1304$-$02           & 13\,04\,46.6&$-$02\,52\,16 && 1122 &   9 && 16.81 & $-$14.23 &            &        \\[2.0mm]

J1310+34   & A & UGC\,8246                    & 13\,10\,04.5&$+$34\,10\,51 &&  833 &   5 && 14.55 & $-$15.78 & NGC\,5005  &  1.38  \\
           & B & SDSS\,J131029.12+341411.5    & 13\,10\,29.2&$+$34\,14\,13 &&  873 &  75 && 18.92 & $-$11.51 &            &        \\[2.0mm]

J1315+47   & A & DDO\,169\,NW                 & 13\,15\,18.4&$+$47\,32\,00 &&  328 &  75 &&       &          & NGC\,4736  &  1.57  \\
           & B & UGC\,08331                   & 13\,15\,30.3&$+$47\,29\,56 &&  344 &   6 && 14.60 & $-$13.82 &            &        \\[2.0mm]

J1337+32   & A & SDSS\,J133605.53+320823.2    & 13\,36\,05.6&$+$32\,08\,21 &&      &     && 17.94 &          & NGC\,5353  & 36.57  \\
           & B & UGC\,8602                    & 13\,36\,45.5&$+$32\,05\,28 && 3062 &   5 && 17.54 & $-$15.63 &            &        \\
           & C & UGC\,8605                    & 13\,36\,54.3&$+$32\,05\,44 && 3035 &   5 && 17.57 & $-$15.58 &            &        \\
           & D & SDSS\,J133657.55+320208.4    & 13\,36\,57.5&$+$32\,02\,08 && 3106 &  29 && 16.90 & $-$16.30 &            &        \\
           & E & UGC\,8608                    & 13\,37\,00.9&$+$31\,46\,00 && 3033 &   3 && 16.04 & $-$17.13 &            &        \\
           & F & SDSS\,J133704.69+315337.9    & 13\,37\,04.7&$+$31\,53\,37 &&      &     && 18.23 &          &            &         \\[2.0mm]

J1355+04   & A & KKH\,86                      & 13\,54\,33.5&$+$04\,14\,35 &&  209 &   5 && 16.88 & $-$9.36  & LG         &  9.56  \\
           & B & SDSS\,J135429.53+041237.1    & 13\,54\,29.5&$+$04\,12\,37 &&      &       &&       &          &            &        \\[2.0mm]

J1404+61   & A & UGC\,08982                   & 14\,03\,00.0&$+$61\,45\,04 && 1868 &  31 && 15.48 & $-$16.62 & NGC\,5322  &  1.02  \\
           & B & SDSS\,J140524.63+613401.3    & 14\,05\,24.6&$+$61\,34\,01 && 1883 &   5 && 16.81 & $-$15.31 &            &        \\[2.0mm]

J1423+21   & A & SDSS\,J142332.69+213112.1    & 14\,23\,32.8&$+$21\,31\,18 &&      &     && 19.86 &          & NGC\,4472  & 50.1   \\
           & B & SDSS\,J142337.04+213128.2    & 14\,23\,37.0&$+$21\,31\,28 && 2068 &  21 && 19.12 & $-$13.31 &            &        \\[2.0mm]

J1428+21   & A & UGC\,9274                    & 14\,28\,02.8&$+$21\,18\,14 && 1117 &  73 && 14.35 & $-$16.69 & NGC\,4472  & 32.34  \\
           & B & UGC\,9282                    & 14\,28\,41.6&$+$21\,20\,22 && 1155 &   5 && 16.17 & $-$14.96 &            &        \\[2.0mm]

J1437+59   & A & SDSS\,J143659.37+590535.1    & 14\,36\,59.3&$+$59\,05\,36 &&      &     && 18.34 &          & NGC\,5777  &  6.70  \\
           & B & SDSS\,J143703.11+590606.3    & 14\,37\,02.0&$+$59\,05\,55 && 2402 &  10 && 18.34 & $-$14.28 &            &        \\[2.0mm]

J1450+36   & A & SDSS\,J144948.75+362347.3    & 14\,49\,48.8&$+$36\,23\,47 && 1979 &   5 && 16.96 & $-$15.26 & UGC\,9519  & 69.09  \\
           & B & SDSS\,J144951.10+362501.5    & 14\,49\,51.1&$+$36\,25\,02 && 1978 &  32 && 16.69 & $-$15.53 &            &        \\[2.0mm]

J1648+21   & A & SDSS\,J164711.10+210527.0    & 16\,47\,11.1&$+$21\,05\,27 &&2727& 2 &$^{\dag}$& 16.76 & $-$16.32 & NGC\,6181  & 30.25  \\
           & B & UGC\,10549                   & 16\,47\,26.5&$+$21\,07\,32 && 2719 &   9 && 16.23 & $-$16.87 &            &        \\
           & C & SDSS\,J164802.08+213330.5    & 16\,48\,02.1&$+$21\,33\,30 &&2753&12 &$^{\dag}$& 17.48 & $-$15.71 &            &        \\[2.0mm]

J1657+38   & A & UGC\,10625                   & 16\,57\,23.1&$+$38\,40\,19 && 2252 &   6 && 16.88 & $-$15.63 & NGC\,6339  &370.21  \\
           & B & SHOC\,553                    & 16\,57\,30.0&$+$38\,41\,23 && 2255 &   5 && 16.90 & $-$15.62 &            &        \\[2.0mm]

J2227$-$09 & A & 6dF\,J2227305$-$093959       & 22\,27\,30.5&$-$09\,39\,59 && 1866 &  74 && 15.64 & $-$16.62 & NGC\,253   &313.57  \\
           & B & MCG\,$-$02-57-007            & 22\,27\,41.5&$-$09\,43\,37 && 1786 &  43 && 15.81 & $-$16.35 &            &
\end{longtable*}

\twocolumngrid



While compiling our list of dwarf galaxy groups we found that some
of the SDSS redshifts were determined using the cross-correlation
method. Though the emission lines, sometimes very bright, are
present in the spectrum, for some reason no velocity estimates
were made based on these lines (the SDSS provides no table of
emission-based redshifts). As a result, the accuracy of the
line-of-sight velocity determinations for such objects proved to
be substantially lower. In some other cases, the redshifts were
determined incorrectly. Using the SDSS spectra we measured the
heliocentric velocities from the emission lines of several
galaxies:
\begin{list}{}{
\setlength\leftmargin{0in}
\setlength\parsep{0in}
\setlength\itemsep{0in} }
\item J0817+24B = KUG\,0814+251,
\item {J0959+41B = KUG\,0956+419},
\item {J1648+21A = SDSS\,J164711.10+210527.0},  and
\item {J1648+21C = SDSS\,J164802.08+213330.5}.
\end{list}
The velocity values of these galaxies are marked by the $^{\dag}$
sign.

\textbf{J0714+44}. The distances to the galaxies of this very
isolated group were determined by Makarova and
Karachentsev~\cite{makarova1998:Makarov} by the brightest blue
stars. The two galaxies are located at the same distance within
the measurement errors, 7.2~Mpc and 7.9~Mpc for UGC\,3698 and
NGC\,2337, respectively, and 8.6~Mpc for UGC\,3817 with a typical
error of around \mbox{$(1\div1.5)$ Mpc}. This also provides yet
another independent evidence of the physical boundedness of
galaxies in the system. Note that NGC\,2337 and UGC\,3698 are
located at a projected distance of  25~kpc, whereas UGC\,3817
resides $2\fdg3$ from them or at a projected distance of 0.3~Mpc.
Such separations are typical for associations of dwarf galaxies.

\textbf{J0742+16}. This pair, together with UGC\,3755 and
UGC\,4115, forms the  14+19 association from Tully et
al.~\cite{tully06:Makarov}. All these galaxies have precise
distance measurements based on the tip of the red giant branch
method. Despite their small line-of-sight velocities, they are
located at the distances of 8.05~(UGC\,3974) and \mbox{8.02
(CGCG\,087-033) Mpc}. A high peculiar velocity of around
$-420$~km/s is due to the ``local velocity
anomaly''~\cite{tully92:Makarov}.

\textbf{J0911+42} structurally resembles the nearby bright
irregular galaxy NGC\,4449 with ongoing star formation, which is
merging the  dwarf low surface brightness  spheroidal galaxy
d1228+4358~\cite{kkh2007:Makarov,hkk2009:Makarov} in what is
believed to be a case of a dwarf galaxy merger
in~\cite{md2011:Makarov}. Note that the J0911+42 pair is much more
isolated and its primary galaxy is about 3~mag fainter than
NGC\,4449 (note that  NGC\,4449 was not included in our list due
to its luminosity \mbox{$M_K=-20.4$}).

\textbf{J1110+40\,B} is a low surface brightness dwarf galaxy with
an unknown line-of-sight velocity, which is located at a distance
of $1\farcm9$ (at a projected distance of about 23~kpc) south of
KUG\,1107+403 and is a possible companion of this galaxy.

\textbf{J1131$-$35} is a close pair of galaxies with a separation
of $16^{\prime\prime}$ corresponding to a projected mutual
distance of 2.6~kpc from each other. Velocity is known only for
one of these galaxies.

\textbf{J1150$-$00}. At an angular distance of $3^\prime$
northwest of the UM\,456 and UM\,456A pair the UM\,455 galaxy is
located, but  it has a substantially higher \linebreak
line-of-sight velocity (\mbox{$V_{\rm LG}=3680$~km/s}).

\textbf{J1244+62} is a chain of five dwarf galaxies extending over
180~kpc. The system appears to be in the process of formation.

\textbf{J1310+34} The disk of the UGC\,8246 galaxy is warped,
possibly, due to the interaction with the second component of the
pair.

\textbf{J1337+32} is one of the most populated groups in our list.
It contains six members. One of the galaxies, J1337+32\,F was
included into the group based on the morphological criteria.
According to the SDSS, the velocity of this galaxy,
SDSS\,J133704.69+315337.9 is equal to \mbox{$V_h=41791$~km/s}.
However, the quality of the spectrum does not allow to reliably
estimate its redshift. The morphology of this low surface
brightness galaxy suggests that the velocity estimate in question
is erroneous.

\begin{figure*}
\setcaptionmargin{0mm}
\includegraphics[scale=0.19]{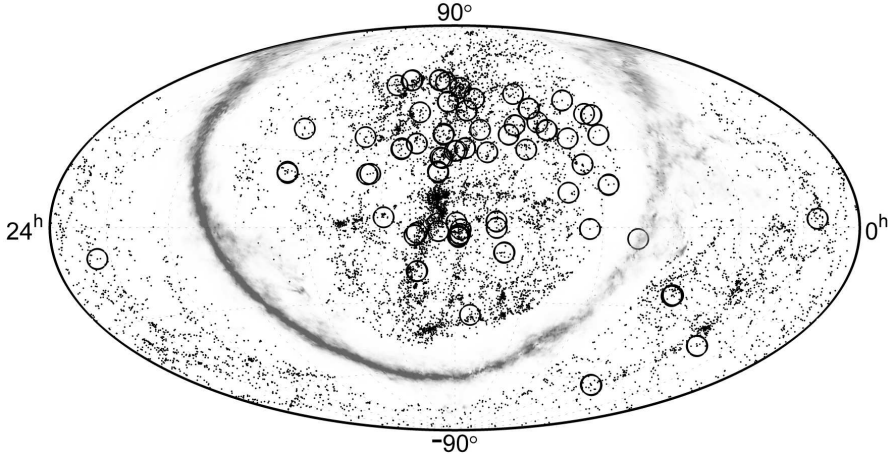}
\caption{The distribution of groups of dwarf galaxies on the sky.
The dots indicate the positions of nearby (\mbox{$V_{\rm
LG}<3500$~km/s}) galaxies. Groups of galaxies are marked by
circles. The Milky Way is shown by gray
clouds.}\label{sky:Makarov}
\end{figure*}

\textbf{J1355+04\,B} was discovered by I.~Karachentsev (private
communication) at an angular distance of $2\farcm2$ (corresponding
to a projected distance of 1.6~kpc) from the nearby isolated dwarf
galaxy KKH\,98. J1355+04\,B is 2--2.5~mag fainter than KKH\,98,
which, in the case if the two galaxies are located at the same
distance, makes it comparable in brightness to the ultra-faint
dwarfs in our Galaxy.

\textbf{J1423+21} is a pair of dwarf galaxies with an angular
separation of $1^{\prime}$ corresponding to a projected separation
of 8~kpc. The line-of-sight velocity \mbox{$V_h=2049$} of the
ADBS\,142335+2131 object was measured in the process of the
Arecibo Dual-Beam Survey~\cite{ADBS:Makarov} and may refer to both
objects.

\textbf{J1437+59} is a close pair of dwarf galaxies with an angular
separation of  $43^{\prime\prime}$ corresponding to a projected
separation of 7~kpc. Velocity is known only for one object.

\textbf{J1648+21}. The complex structure of the \linebreak
J1648+21\,A galaxy is either due  to a burst of star formation or
to the tidal influence of the UGC\,10549 galaxy despite their
rather large separation  (the projected separation is equal to
45~kpc). The third member of the group, J1648+21C, is located
$29^\prime$ north of the two brighter galaxies at a projected
distance of 300~kpc.

\section{MAIN PROPERTIES OF DWARF GALAXY GROUPS}

Figure~\ref{sky:Makarov} shows the sky distribution of groups of
dwarf galaxies. The overwhelming majority of the systems are
concentrated in one third of the sky covered by the  SDSS survey.
This fact reflects strong observational selection, inherent to our
sample. Unfortunately, the selection effects are practically
impossible to account for. The factors affecting the completeness
of our data include both the coverage of modern sky surveys and
selection of candidates for the further spectroscopic analysis.
Low-luminosity and low-surface brightness galaxies are usually
unobservable. One may expect the real number of groups consisting
of dwarf galaxies alone to be appreciably greater than the \gdN{}
groups already identified (about 5\% of the total number of
groups) on the scale of the Local Supercluster. Despite the
selection effects, it can be pointed out that multiple dwarf
systems avoid the known concentrations of luminous matter, and
their distribution is substantially more uniform. Note that dwarf
galaxies evade the \mbox{$15^h<\alpha<18^h$} region. This is
especially evident from a comparison with a similar region located
symmetrically with respect to the plane of the Local Supercluster.
This fact appears to be due to an extreme deficit of dwarf
galaxies in the Local Void~\cite{LocalVoid:Makarov}.

\begin{figure}
\setcaptionmargin{0mm}
\includegraphics[width=\columnwidth,clip]{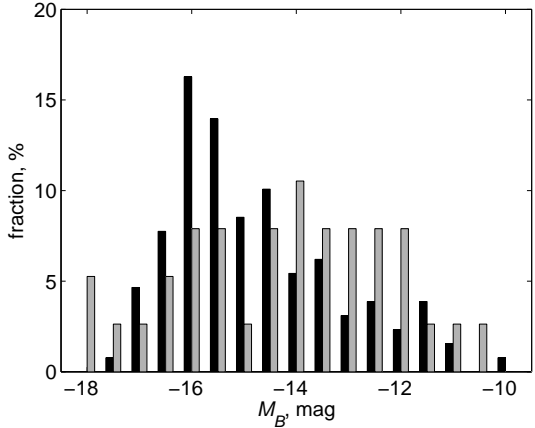}
\caption{The luminosity function of galaxies of our sample
compared to the associations of Tully et
al.~\cite{tully06:Makarov}. } \label{f:lf:Makarov}
\end{figure}

Figure~\ref{f:lf:Makarov} shows the luminosity function of the
objects of our sample compared to associations of dwarf galaxies.
The groups of dwarf galaxies (shown in black) occupy the same
interval of absolute magnitudes as associations (shown in light
gray). The completeness of data of the Catalog of Neighboring
Galaxies (CNG)~\cite{CNG:Makarov} approaches 100\% on a 2~Mpc
scale down to an absolute magnitude of \mbox{$M_B=-10$}. It is
evident that the sharp decrease of the number of galaxies in our
list with luminosities below \mbox{$M_B = -15\fm5$} is due to the
observational selection. Let us assume that the luminosity
function of galaxies in groups of dwarfs is proportional to the
distribution obtained for galaxies located out to 3~Mpc, and take
into account the fact that the SDSS survey, which is the main
contributor to our data, covers about one third of the sky. We can
hence make a crude estimate of the total number of systems of
dwarf galaxies on the scale of the Local Supercluster. There
should be five to six times more groups of dwarf galaxies than we
have actually found so far.

\begin{figure}
\setcaptionmargin{0mm}
\includegraphics[width=\columnwidth,clip]{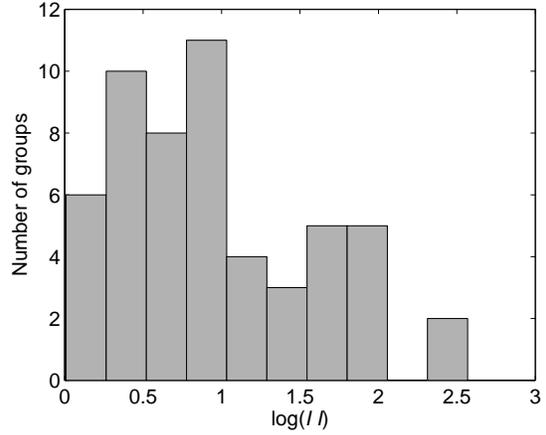}
\caption{The distribution of isolation indices of the dwarf galaxy
groups.} \label{f:ii:Makarov}
\end{figure}

It is evident from Fig.~\ref{f:ii:Makarov} that the groups of
dwarf galaxies span over a wide range of values of the isolation
index ($II$). This index is equal to the factor by which the
masses of all galaxies should be multiplied for the given group or
galaxy to be gravitationally bound with other systems.
\mbox{$II\sim1$} means that the system in question is located near
the zone of gravitational influence or, in other words, is near
the zero-velocity surface of its more massive companion. Large
$II$ values correspond to the regions of low matter density in the
Universe, which are located far from massive gravitating centers.

\begin{figure}[t]
\setcaptionmargin{0mm}
\includegraphics[width=\columnwidth,clip]{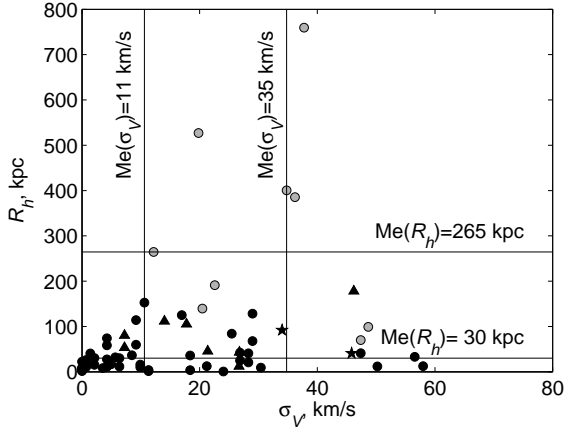}
\caption{The dependence of the size of groups from the internal
velocity dispersion. The black symbols show the groups of our
sample (the circles, triangles, and asterisks correspond to pairs,
triplets, and more populated systems, respectively). The gray
circles show the associations of Tully et
al.~\cite{tully06:Makarov}. } \label{f:rsig:Makarov}
\end{figure}

Figure~\ref{f:rsig:Makarov} shows the distribution of groups by
the size and internal motions within the system. The black symbols
indicate the groups of dwarfs with different numbers of members:
the dots, triangles, and asterisks show pairs, triplets, and more
populated groups, respectively. The dispersion of the
line-of-sight velocities ($\sigma_V$) in groups of dwarfs reaches
60~km/s with a median value of \gdVelDisp~km/s, and the projected
sizes of groups do not exceed  200~kpc with the median value
amounting to \gdRh{}~kpc.

\begin{table}[b]
\setcaptionmargin{0mm} \onelinecaptionsfalse \captionstyle{normal}
\caption{A comparison of the main parameters of normal groups of
the Local Supercluster (NGLC), groups of dwarf galaxies (GD), and
associations of dwarf galaxies (AD)}
\begin{tabular}{l|r|c|c|c|c|c}
\hline
 & \multicolumn{1}{c|}{{$n$}} & $\sigma_{V}$, & $R_h$, & $M_p$,  & $L$, & $M/L $, \\
 &   & km/s & kpc & $10^{10}~M_{\sun}$ & $10^9~L_{\sun}$ & $M_{\sun}/ L_{\sun} $ \\
\hline
 NGLC                                       & 1082 & 42 & 160 & 61 & 42  & 21 \\
\multicolumn{1}{r|}{$n=2$}         &   516 &  24 & 121 & 14 & 17 &  11 \\
\multicolumn{1}{r|}{$n=3$}          &  171 & 41 & 156 &  46 & 40 &  15 \\
\multicolumn{1}{r|}{$n\ge4$}    &  395 & 74 & 204 & 330& 120 & 31 \\
\hline
 AD & 7    & 35 & 265 & 38 & 1.0 & 380 \\
\hline
 GD                                              & \gdN  & \gdVelDisp & \gdRh &
0.96 & 0.35 & \gdML \\
\multicolumn{1}{r|}{$n=2$}          & 47 &  9 & 22 &  0.29  & 0.29 &  26 \\
\multicolumn{1}{r|}{$n=3$}           &  8 & 20 & 67 &  4.6 & 0.69 &  83 \\
\multicolumn{1}{r|}{$n\ge4$}     &  2 & 40 & 66 &  26  & 2.0 & 129 \\
\hline
\end{tabular}
\label{t:comparison:Makarov}
\end{table}

\begin{figure*}
\setcaptionmargin{5mm} \onelinecaptionsfalse
\includegraphics[scale=0.5,clip]{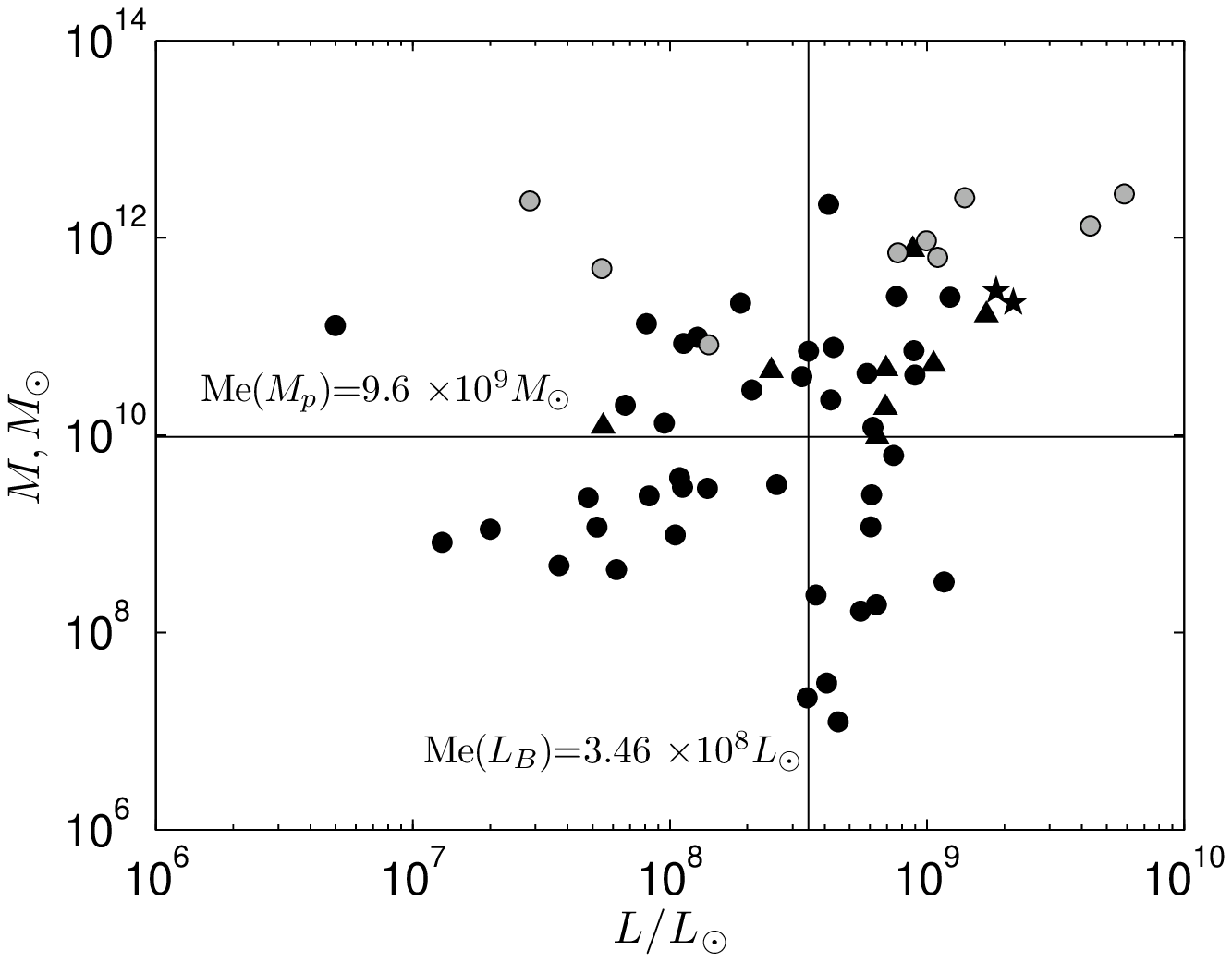}
\hspace{2mm}
\includegraphics[scale=0.5,clip]{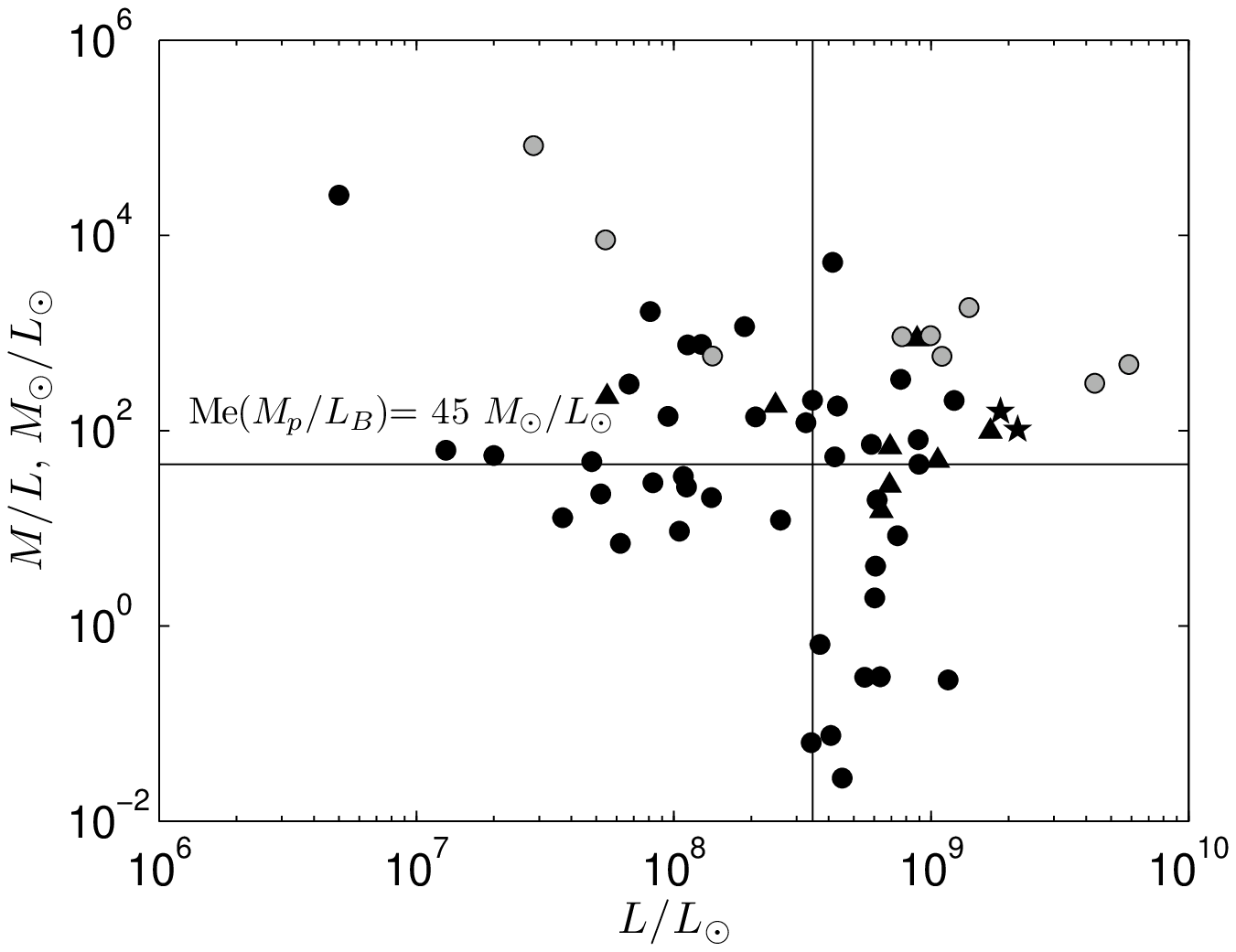}
\captionstyle{normal} \caption{The left-hand panel demonstrates
the mass-to-luminosity relation for groups and associations of
dwarf galaxies; the right-hand panel shows the
``'mass-to-luminosity ratio'--luminosity'' relation. The black
symbols mark the groups of our sample (the dots, triangles, and
asterisks show the pairs, triplets, and groups with  $n\ge4$
members, respectively). The gray circles show the associations of
dwarf galaxies. } \label{f:ml:Makarov}
\end{figure*}

Table~\ref{t:comparison:Makarov} summarizes the main parameters of
groups of galaxies in the Local Supercluster, the associations and
groups of dwarf galaxies. This table gives the median values of
velocity dispersion $\sigma_V$, the harmonic radius of systems
$R_h$, projected mass $M_p$, luminosity $L$, and
mass-to-luminosity ratio $M/L$. For the groups in the Local
Supercluster and the groups of dwarfs the parameters are given
both for the entire sample of these groups and for particular
configurations: pairs \mbox{($n=2$)}, triplets \mbox{($n=3$)}, and
other groups \mbox{($n\ge4$)}. A comparison of the data shows that
the groups of dwarfs are the most compact formations. The sizes of
groups from our sample are about one order of magnitude smaller
than those of normal groups and associations. Velocity dispersions
do not differ so drastically, however, groups of dwarf galaxies
again possess the lowest values, which are a factor of seven and
three smaller than those of normal groups, and associations,
respectively. Nevertheless, the dwarf systems are  intermediate
between groups of galaxies and associations in terms of the
mass-to-luminosity ratio. The list of dwarf groups consists of
47~pairs, 8~triplets, and only two more populated groups. This
statistics is too small to analyze the differences between the
systems with different numbers of neighbors. Note, however, that
the physical parameters  systematically vary with the growing
number of group members. The triplets of dwarf galaxies prove to
be systematically wider, more massive, and to contain more
luminous matter compared to similar pairs of dwarfs. The
mass-to-luminosity ratio also increases with the number of members
in the group.

The gray circles in Fig.~\ref{f:rsig:Makarov} show the
distribution of associations of dwarf
galaxies~\cite{tully06:Makarov}. The associations  were discovered
in the Local Volume exclusively via analyzing the spatial
distribution of galaxies based on high-precision distance
measurements. As Tully et al.~\cite{tully06:Makarov} pointed out,
almost all dwarf galaxies in the Local Volume, except for KKR\,25,
are the members of groups and associations. The associations are
rather sparse structures. A typical size of an association
(\mbox{$R_h=265$~kpc}) is almost equal to the typical size of
groups of normal galaxies, whereas the total luminosity of
associations is two orders of magnitude lower than that of groups
in the Local Supercluster. Although the groups and associations
differ by a factor of three in terms of velocity dispersions, they
differ substantially, by a factor of nine, in terms of the
characteristic size. This fact is a manifestation of the
fundamental difference between the sample construction processes.
The associations were identified based on the spatial correlation
of objects, whereas the groups of galaxies were identified based
on the kinematical data only, selecting candidates for physically
bound groups of galaxies. The requirement that the groups should
be gravitationally bound, combined with the luminosity-based mass
estimates, necessarily implies small projected distances and small
velocity differences between the companions of dwarf systems.

The median \mbox{$B$-band} luminosity of groups of dwarf galaxies
is \gdLum{}~$L_{\sun}$, and the median projected mass value is
\mbox{\gdMass{}~$M_{\sun}$}. This implies a mass-to-luminosity
ratio of \gdML{}~$M_{\sun}/L_{\sun}$. Note that individual mass
estimates for groups of dwarf galaxies are highly uncertain
because of the small multiplicity of the systems (which are mostly
binary galaxies). We can therefore deal with ensemble-averaged
quantities. Figure~\ref{f:ml:Makarov} shows the dependences of the
mass and the  mass-to-luminosity ratio on the total luminosity. It
is evident from the figure that the associations are, on the
average, more massive than the groups of dwarfs. It should,
however, be noted that despite the use of different identification
algorithms and substantial difference in their sizes and velocity
dispersions, groups and associations of dwarf galaxies form a
continuous sequence on the ``mass--luminosity'' diagrams. This
fact is a manifestation of the genetic relationship of these
systems.

\section{CONCLUSIONS}

Over the past decade modern mass surveys have substantially
increased the number of galaxies with known velocities in the
Local Supercluster. We constructed a catalog of systems consisting
of dwarf galaxies only based on the catalog of groups in the Local
Supercluster~\cite{makarov11_groups:Makarov}. Our catalog contains
groups where the brightest galaxy has a $K$-band luminosity lower
than \mbox{$M_K=-19$}. Such systems make up about 5\% of all
groups in the Local Supercluster. However, with selection effects
taken into account, the total number of multiple dwarf systems
should be at least a factor of five to six greater. The groups of
dwarf galaxies are characterized by the mean size of \gdRh{}~kpc
and a mean velocity dispersion of \gdVelDisp{}~km/s. Both these
values are much smaller than the corresponding parameters for
typical groups in the Local Supercluster (204~kpc and 74~km/s,
respectively). Our sample of dwarf galaxy groups forms a
continuous sequence in the mass and luminosity distribution
diagrams along with associations identified by Tully et
al.~\cite{tully06:Makarov} based on an analysis of the
three-dimensional distribution of nearby dwarf galaxies. The
groups and associations of dwarfs have similar luminosities,
however, the groups are by one order of magnitude more compact.
The median mass-to-luminosity ratio for the groups of dwarfs is
equal to \mbox{\gdML{} $M_{\sun}/L_{\sun}$}, which is indicative
of a greater amount of dark matter, as compared to normal groups.

The systems of dwarf galaxies may contain substantial amounts of
dark matter. Such ``dark'' aggregates may be quite numerous. They
are difficult to reveal and study and can therefore ``hide'' a
substantial fraction of dark matter, which remains undiscovered in
the studies of groups of galaxies. This may partially solve the
problem of ``missing''  \linebreak mass---the discrepancy between
the estimates of the average density of the Universe based on the
analyses of cosmic background radiation, and the estimates ensuing
from the analysis of groups of galaxies in the Local Supercluster
\cite{makarov11_groups:Makarov}.

Note that the issues of the formation and evolution of the systems
of dwarf galaxies remain highly unexplored. This is due to the
difficulties one has to face in the process of observations and
interpretation, and the problems of the theoretical approach. When
we study groups consisting of dwarf galaxies only, we have to
address a large number of problems: the low surface brightness and
low luminosity make such systems     hardly accessible for
observations, whereas their small masses impose very severe
constraints in cosmological computations. Our list of dwarf galaxy
groups formed the basis for the spectroscopic survey, currently
conducted at the 6-m telescope of the Special Astrophysical
Observatory of the Russian Academy of Sciences since 2008. The
main goal of the survey is to study the chemical composition of
dwarf galaxies in groups and determine the evolutionary status of
such systems.

\begin{acknowledgments}
This work was supported by the Russian Foundation for Basic
Research (grant nos.~11--02--00639 and 11--02--90449) and  the
Ministry of Education and Science of Russian Federation (state
contracts no.~14.740.11.0901, 16.552.11.7028, 16.518.11.7073). The
research has made use of the HyperLEDA database ({\tt
http://leda.univ-lyon1.fr}).
\end{acknowledgments}


\end{document}